\documentclass[12pt]{article}

\usepackage{graphicx}
\usepackage{booktabs}
\usepackage{ragged2e} 
\usepackage{supertabular}
\usepackage[dvipsnames]{xcolor}
\usepackage{caption}
\usepackage{subcaption}
\usepackage{float}
\usepackage{mathtools}
\definecolor{nblue}{RGB}{28,130,185}
\definecolor{cgreen}{RGB}{76,153,0}
\definecolor{myorange}{RGB}{245,156,74}

\usepackage{hyperref}
\hypersetup{
  colorlinks=true,
  citecolor=magenta,
  urlcolor=-myorange
}

\usepackage{amssymb}
\usepackage{amsmath}
\usepackage{amsfonts}
\newcommand{\be}{\begin{equation}}
\newcommand{\ee}{\end{equation}}
\newcommand{\bea}{\begin{eqnarray}}
\newcommand{\eea}{\end{eqnarray}}
\newcommand{\nn}{\nonumber}

\begin{document}

\title{The Photon Gas in Classical Mechanics: A Statistical-Mechanical Treatment of Classical Field Theory}

\author{Farhang Loran\thanks{E-mail address:
loran@iut.ac.ir} ~ and Saman Moghimi-Araghi\thanks{E-mail address: samanimi@sharif.edu }
\\[6pt]
$^{*}$Department of Physics, Isfahan University of Technology,\\  Isfahan 84156-83111, Iran\\
$^\dagger$Department of Physics, Sharif University of Technology, \\ Tehran, P.O. Box:11555-9161, Iran \\[6pt]
}

\date{ }
\maketitle

\begin{abstract}
The thermodynamics of cavity radiation is usually introduced through the quantum theory of a photon gas. In contrast, we investigate which features of thermal radiation follow from classical electrodynamics alone. We show, using two complementary approaches, that the proportionalities between energy density, radiation pressure, and energy flux arise without invoking photons or quantum statistics. The first approach is based on explicit solutions of Maxwell's equations and spatio-temporal averaging of electromagnetic waves. The second approach develops the classical statistical mechanics of the electromagnetic field from first principles by constructing the canonical Hamiltonian, implementing the gauge constraints, formulating the partition function as a functional integral, and evaluating the corresponding field correlation functions. By combining wave analysis, statistical mechanics, constrained Hamiltonian dynamics, and functional methods within a single classical framework, this work provides a coherent classical formulation of electromagnetic radiation thermodynamics.
\vspace{3mm}

\end{abstract}
\section{Introduction}

Thermodynamics and statistical mechanics of electromagnetic radiation confined in a cavity is usually presented as the problem of a photon gas \cite{Harvey S. Leff}. The photon picture -- massless bosons obeying Bose-Einstein statistics -- gives a clean and straightforward derivation of the Planck spectrum, the Stefan-Boltzmann law, and the relations between energy density, radiation pressure, and energy flux. This quantum approach is so successful that one might think little else needs to be said.

Historically, however, important results concerning thermal radiation were obtained before the introduction of the light quantum in 1905 \cite{Britannica,history}. The Stefan-Boltzmann law was first discovered experimentally by Stefan in 1879 and then derived theoretically by Boltzmann in 1884 \cite{Boltzmann84}. Boltzmann's derivation used only thermodynamics and the relation $P = u/3$ between the pressure $P$ and the energy density $u$ -- the latter following from Maxwell's electrodynamics. But there is a subtlety: Boltzmann's argument implicitly assumed that the internal energy $U$ is a finite function of temperature and volume, an assumption that classical electrodynamics could not justify. Indeed, when one attempts a purely classical calculation of the energy density by counting modes and applying the equipartition theorem, one runs into the ultraviolet catastrophe and a divergent result. Thus, while the Stefan-Boltzmann law was \emph{obtained} before quantum theory, it was not \emph{derivable} from first principles within a consistent classical framework without additional (and ultimately unjustified) assumptions.

Given this failure, one might ask: why revisit the classical treatment at all, let alone in a systematic field-theoretic manner? The answer is twofold. First, despite its inability to produce the correct spectrum or the correct absolute value of the Stefan-Boltzmann constant, classical electromagnetism can still predict certain universal \emph{relations} between observable quantities correctly -- in particular, the proportionality between energy flux, $I$ and energy density, $u$, and similarly the proportionality between radiation pressure, $P$, and energy density. These proportionalities, when combined with standard thermodynamics, yield the Stefan-Boltzmann form $u \propto T^4$; only the numerical constant remains undetermined (and, in the classical calculation, divergent). Second, and more importantly for the present paper, electrodynamics is a classical gauge field theory. Constructing its statistical mechanics from scratch -- starting from the Lagrangian, identifying the constraints, fixing the gauge, writing the partition function as a functional integral, and computing thermal averages --- is a nontrivial exercise that rarely appears in textbooks on thermal physics. This exercise provides a valuable playground for understanding how field theory, gauge invariance, and statistical mechanics interact, and it serves as a natural warm-up for finite-temperature quantum field theory and path integral methods. Other attempts to derive the blackbody spectrum within a purely classical framework, invoking assumptions such as zero‑point radiation or stochastic electrodynamics, have also been explored in the literature; see, for example, the works of Boyer and Cole \cite{Boyer1989,Cole1990,Cole1992,Cole2000}.

Numerous treatments of electromagnetic radiation discuss individual aspects of radiation pressure, blackbody thermodynamics, or the Hamiltonian formulation of Maxwell theory. The present work differs in that it presents these topics within a single self-contained classical framework that combines wave analysis, statistical mechanics, constrained Hamiltonian dynamics, and functional methods. This unified presentation is intended to provide advanced students with a coherent bridge between classical electrodynamics and finite-temperature field theory. 

In this paper, we focus on what the classical field theory can do. We present two complementary classical derivations of the proportionalities $I \propto u$ and $P \propto u$. The first, is more intuitive: it solves Maxwell's equations, decomposes the fields into Fourier modes, and uses spatial and temporal averaging to obtain the desired relations. The second, is the methodological core of the paper: we start from the Lagrangian, move to the Hamiltonian formalism, explicitly handle the gauge freedom, write the classical partition function as a functional integral, compute field correlation functions, and finally derive $P = u/3$ from the Maxwell stress tensor. Along the way, we also see why the average Poynting vector vanishes in equilibrium, a fact that distinguishes the equilibrium ensemble approach from the first wave-based derivation, where modes propagating in opposite directions can be treated separately. 

The paper is organized as follows. Section~\ref{sec:photon} briefly recalls the standard photon-gas argument for context,  and show how the proportionalities lead to the Stefan-Boltzmann law via classical thermodynamics. Section~\ref{sec:intuitive} gives the intuitive classical wave-based derivation. Section~\ref{sec:formalism} contains the systematic, constraint-based, field-theoretic treatment. We conclude in Section~\ref{sec:conclusion} with a discussion of what classical physics can and cannot achieve in this problem, and why the exercise remains useful for understanding the interface between field theory, statistical mechanics, and gauge invariance. Technical details and supporting derivations are provided in the Appendices.

\section{Photon gas and thermodynamic relations}
\label{sec:photon}

\begin{figure}
  \centering
  \includegraphics[width=0.5\textwidth]{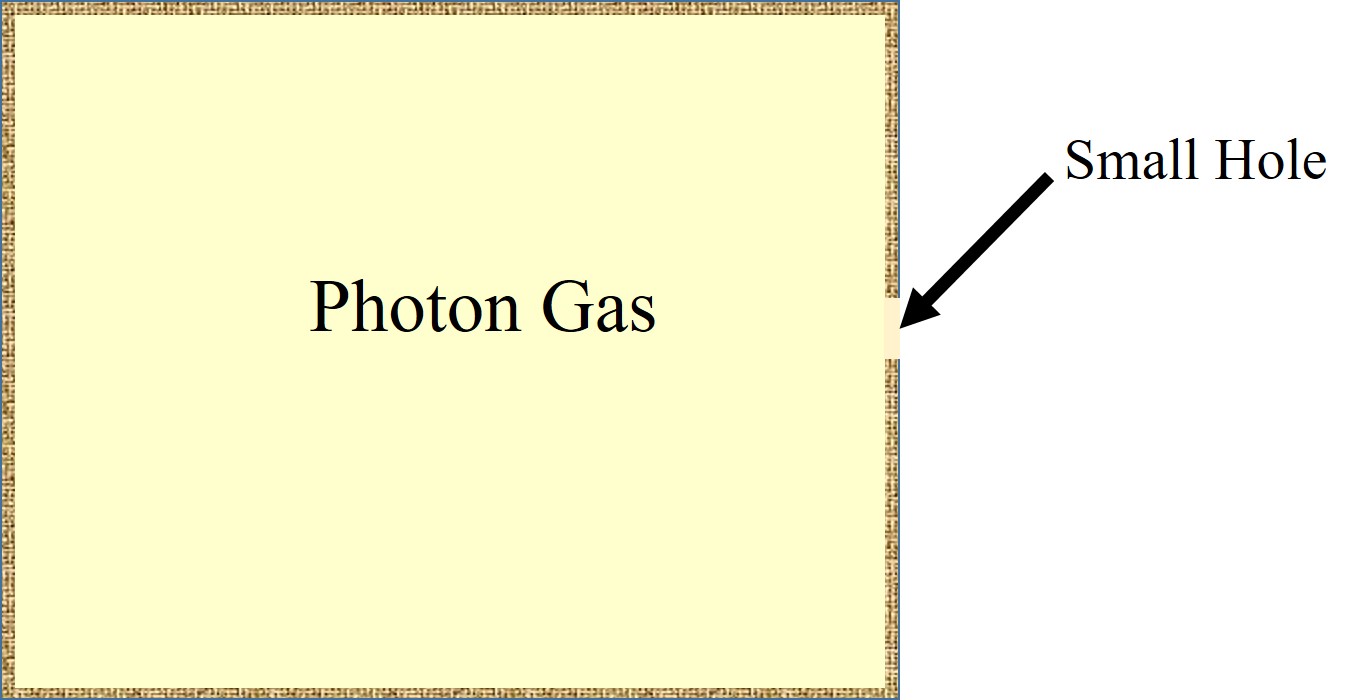}
  \caption{A cavity filled with a photon gas and held at temperature \(T\). The small hole acts as a blackbody aperture through which the equilibrium radiation can escape and be observed.}\label{Fig-1}
\end{figure}

Consider a photon gas consisting of photons confined within a cavity, as illustrated in Figure~\ref{Fig-1}, and maintained in thermal equilibrium with the cavity walls at temperature \(T\). In equilibrium, the energy radiated per unit surface area per unit time obeys the Stefan-Boltzmann law, i.e., it is proportional to the fourth power of the temperature. In addition to this law, one also can obtain that the energy density is proportional both to the rate of energy emission per unit area and to the radiation pressure.

Remarkably, the Stefan-Boltzmann law can be obtained without any reference to the quantum nature of light, using only classical thermodynamics together with the relation between radiation pressure and energy density \cite{Harvey S. Leff 2002}. Suppose we know from electrodynamics (or from experiment) that for isotropic radiation
\begin{equation}
P = \frac{u}{3},    
\end{equation}
where $P$ is the radiation pressure and $u \coloneqq U/V$ is the energy density. The first law of thermodynamics,
\begin{equation}
dU = TdS - PdV,    
\end{equation}
gives, at constant temperature,
\begin{equation}
\left(\frac{\partial U}{\partial V}\right)_T = T\left(\frac{\partial S}{\partial V}\right)_T - P.
\end{equation}
Using the Maxwell relation that follows from the Helmholtz free energy \cite{Maxwell},
\begin{equation}
\left(\frac{\partial S}{\partial V}\right)_T = \left(\frac{\partial P}{\partial T}\right)_V,
\end{equation}
we obtain
\begin{equation}
 \left(\frac{\partial U}{\partial V}\right)_T = T\left(\frac{\partial P}{\partial T}\right)_V - P.   
\end{equation}

Since \(U = V u(T)\) and \(P = u(T)/3\), the left-hand side is simply \(u(T)\), while the right-hand side becomes
\begin{equation}
\frac{T}{3}\frac{du}{dT} - \frac{u}{3}.    
\end{equation}
Equating the two sides yields a differential equation for \(u(T)\), whose solutions is 
the Stefan-Boltzmann law,
\begin{equation}
u = A T^{4},    
\end{equation}
where \(A\) is an undetermined constant. Similar thermodynamic analyses of blackbody radiation have been discussed  by Kelly and by Lee \cite{Kelly,Lee}. Thus, without invoking any quantum hypothesis, the temperature dependence \(u \propto T^{4}\) follows solely from thermodynamics and the classical relation \(P = u/3\). The constant \(A\) cannot be fixed by classical physics; deriving its correct value,
\begin{equation}
    A = \frac{\pi^{2}k_{B}^{4}}{15c^{3}\hbar^{3}},
\end{equation}
requires quantum theory and the Planck distribution,  which is in agreement with the experiment \cite{experiment}.

Now we are left with a question: where do the proportionalities \(P \propto u\) and \(I \propto u\) come from? The standard textbook answer uses the photon picture and kinetic theory (see, for example, \cite{Blundell}). Consider a cavity as shown in Figure~\ref{Fig-1}. We aim to determine the intensity of the radiation that escapes through a small hole in the wall. The radiation may contain photons of different frequencies $\omega$. In the photon gas picture employed here, photons are assumed to be non-interacting,\footnote{In quantum electrodynamics (QED), photons can interact with each other via higher order processes such as virtual electron, positron loops \cite{QFT}. These effects are extremely weak at ordinary energies, but they have been experimentally observed \cite{ATLAS}.} so each frequency can be analyzed independently. Thus, we focus on a single frequency at a time.

\begin{figure}
  \centering
  \includegraphics[width=0.5\textwidth]{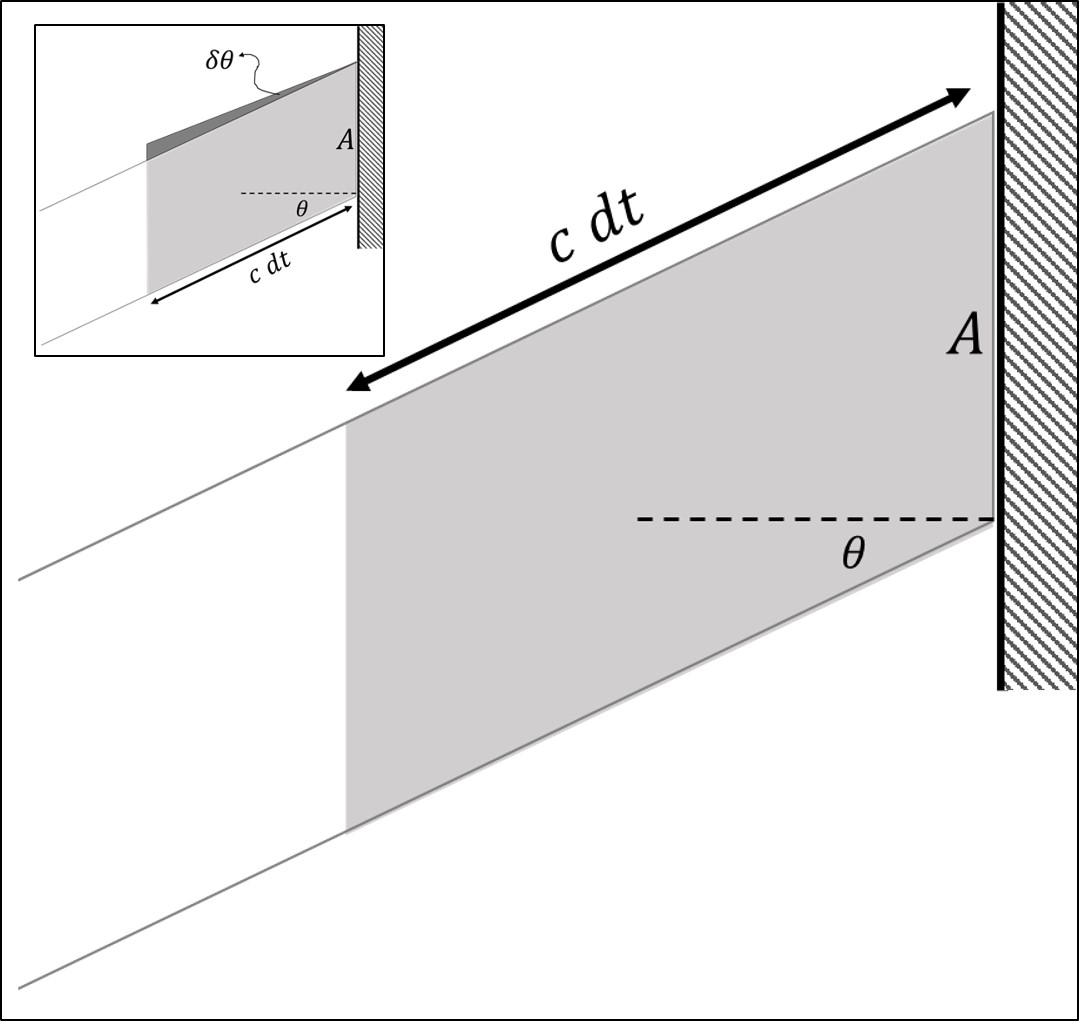}
  \caption{Photons moving at angle $\theta$ reaching a surface element $A$ during the time interval $dt$. The shaded volume corresponds to the photons traveling a distance $c\,dt$. The inset shows the infinitesimal angular spread $\delta\theta$, whose effect on the calculation is negligible to leading order.}\label{Fig-2}
\end{figure}

To calculate the radiation rate, we first determine the number of photons that pass through the small hole in a unit of time. Because the motion of photons is isotropic, the number of photons with velocity within the solid angle $d\Omega$ is independent of the orientation of that angle. In spherical coordinates, the solid angle element is $d\Omega = \sin\theta \, d\theta \, d\varphi$, where $\theta$ is the polar angle (measured from the $z$--axis) and $\varphi$ is the azimuthal angle (measured in the $xy$--plane). If we denote by $f(\theta,\varphi)\,d\Omega$ the probability that photons move in a given direction, then
\[
f(\theta,\varphi) = \frac{1}{4\pi}.
\]
Let us now compute the number of photons that, moving in a given direction, strike a small surface area $A$ of the wall during a short time interval $dt$. Figure~\ref{Fig-2} illustrates the geometry: all photons contained in the parallelogram volume shown in gray, and moving in the specified direction, will reach the surface in time $dt$. One might object that we have drawn only a single angle in the figure, whereas we should consider an interval of angles. This is correct, but if we take an interval such as $\theta$ to $\theta+\delta\theta$, the difference is only that the parallelogram volume is slightly modified, producing a higher-order correction that does not affect the calculation. Thus, the number of photons colliding with the surface is equal to the volume of this parallelogram multiplied by the photon density and by the probability of having such a velocity:
\begin{equation}
N(\omega, \theta, \phi) = (c\, dt \, A \cos \theta) \, n_{\omega} \, \frac{d\Omega}{4\pi},
\end{equation}
where $n_\omega$ is the density of photons with frequency $\omega$.

Two more steps are required to reach the desired result. First, we must integrate over all angles of incidence; second, we must include all frequencies together with their energies. Carrying out the angular integration gives
\begin{equation}
N(\omega) = \int (c\, dt \, A \cos \theta) \, n_{\omega} \, \frac{d\Omega}{4\pi} 
= \frac{c A dt \, n_{\omega}}{4}.
\end{equation}
Note that the integration is taken only over half of the space, corresponding to the side of the wall where the photon gas is present. In this way, the collision flux of photons with frequency $\omega$, i.e., the number of collisions per unit time per unit area, is \(c\, n_\omega/4\).

All that remains is to account for the energy carried by each photon and sum over all frequencies. The energy of a photon is $\hbar\omega$, so the energy flux per unit time per unit area is
\begin{equation}
I = \int d\omega \, \hbar \omega \, \frac{c\, n_{\omega}}{4}.
\end{equation}
The energy density has a similar expression: multiplying the number of photons per unit volume of frequency $\omega$ by their energy and integrating over all frequencies gives
\begin{equation}
u = \int d\omega \, \hbar \omega \, n_{\omega}.
\end{equation}
Comparing the two relations, we immediately find
\begin{equation}
\label{intensity}
I = \frac{c}{4} u.
\end{equation}
This is the result we sought. Similar straightforward arguments also establish the relation between radiation pressure and energy density,
\begin{equation}
\label{pressure}
P = \frac{u}{3},
\end{equation}
which likewise appears in most statistical physics textbooks. In fact, this relation has been obtained from classical reasoning in the seminal book by Landau and Lifshitz for ultra-relativistic particles \cite{LandauLifshitz}.
This derivation, although does not need quantum mechanics, still is based on particle picture of a gas. However, the proportiaonalities should be derivable within the wave picture of classical electromagnetism alone, without assuming the existence of particles or photons. In the following two sections, we present exactly such classical derivations. These derivations show that while the absolute normalization of the Stefan-Boltzmann law requires quantum physics, the functional relations among energy density, flux, and pressure are already contained in classical electromagnetism and classical statistical mechanics.

\section{An Intuitive Approach Based on Classical Electromagnetic Waves}
\label{sec:intuitive}

In thermodynamics we deal with macroscopic quantities. Using statistical mechanics, we try to derive these macroscopic quantities from microscopic ones, usually by some kind of averaging. In real measurements, the averaging is mostly over time and sometimes over a region of space. For example, when we measure the pressure of a car tyre, we use a pressure gauge that has a small but finite area and we leave it in contact with the tyre for a time much longer than the microscopic time scales, then read the pressure. In the theory part, if we wanted to perform time averaging directly, we would normally face the problem that the equations of motion cannot be solved analytically. Therefore we usually turn to ensemble averaging. In the case of electromagnetic waves, however, we are dealing with a system that is exactly solvable analytically: Maxwell's equations in vacuum are linear and their solutions are plane waves. Hence we can follow the more intuitive path in which we start with the direct solution of Maxwell's equations, then perform spatial and temporal averages over scales larger than the microscopic ones (i.e., larger than the typical wavelength and oscillation period), and finally compute the macroscopic quantities of interest. As stated earlier, we focus on the relations between the energy flux, the energy density, and the radiation pressure.

\subsection{Fourier representation of the fields}

From a purely classical standpoint, the description of an empty cavity filled with radiation relies entirely on Maxwell's equations in vacuum, which govern the dynamics of the electric and magnetic fields \(\vec{E}(\vec{x},t)\) and \(\vec{B}(\vec{x},t)\). Moreover, Maxwell's equations ensure that once the electric field is specified, the magnetic field is determined accordingly \cite{Griffiths}. It is therefore sufficient to focus our analysis on \(\vec{E}\).

Maxwell's equations are linear, whose solutions, when expressed in Fourier space decouple and can be analysed individually. In this representation, the dynamics of the system are described by the Fourier components \(\vec{E}(\vec{k})\), where \(\vec{k}\) is the wave vector \cite{Maxwell-Fourier}. The electric field can be written in terms of its Fourier modes as
\begin{equation}
\vec{E}(\vec{x},t) = \int \frac{d^{3}k}{(2\pi)^{3}}\left(\vec{E}(\vec{k})e^{i\vec{k}\cdot \vec{x}}e^{-i\omega (\vec{k})t} + \mathrm{c.c.}\right). 
\label{E-solution}
\end{equation}
Here we have used the same symbol \(\vec{E}\) for both the field and its Fourier components, a common but harmless abuse of notation. Note that the electric field \(\vec{E}(\vec{x},t)\) is real, while its Fourier components \(\vec{E}(\vec{k})\) are complex. {Coulomb's law implies that} 
    \be
    \vec k\cdot\vec E(\vec k)=0.
    \label{coulomb} 
    \ee

From Maxwell's equations, it follows that the dispersion relation is \(\omega(\vec{k}) = c\|\vec{k}\|\), where \(\| \vec{k}\| \coloneqq \sqrt{\vec{k}\cdot\vec{k}}\). {By Faraday's law of induction,} the magnetic field can also be expressed in terms of the Fourier components of the electric field:
    \begin{equation}
    \label{Fourier-magnetic-field}
    \vec{B}(\vec{x},t) = \int \frac{d^{3}k}{(2\pi)^{3}}\left(\frac{\vec{k}}{\omega(\vec{k})}\times \vec{E}(\vec{k})e^{i\vec{k}\cdot \vec{x}}e^{-i\omega (\vec{k})t} + \mathrm{c.c.}\right). 
    \end{equation}

\subsection{Energy density and spatio-temporal averaging}

The specific question we wish to address is the relationship between the energy density and the radiant flux in a chamber filled with electromagnetic fields. We begin by calculating the energy density, which is obtained directly from the fields:
\begin{equation}\label{energy-density-1}
u(\vec{x},t) = \frac{1}{2}\epsilon_0\| \vec{E} (\vec{x},t)\| ^2 +\frac{1}{2\mu_0}\| \vec{B} (\vec{x},t)\| ^2,
\end{equation}
where $\epsilon_0$ and $\mu_0$ denote the vacuum permittivity and permeability, respectively. 

As stated before, in thermodynamics, the energy density is not understood as the instantaneous microscopic quantity introduced above, but rather as an average taken over an appropriate time interval and spatial region. To illustrate this idea, let us recall the analogy with fluid mechanics. When we denote the fluid velocity field by \(\vec{v}(\vec{x},t)\), we do not mean the instantaneous microscopic velocity of a single molecule located at position \(\vec{x}\) at time \(t\). Instead, the velocity field is obtained by averaging the microscopic velocities over a spatial region that is much larger than molecular dimensions yet much smaller than the macroscopic scale of the system. In practice, one also performs an average over a short time interval, long compared to microscopic timescales but short relative to the macroscopic dynamics.

Returning to electromagnetism, we define the thermodynamic energy density as a local spatial average:
\begin{equation}
\underline{u}(\vec{x},t) = \frac{1}{V_{1}(\vec{x})}\int_{V_{1}(\vec{x})}d^{3} x'\,u(\vec{x}^{\prime},t), 
\end{equation}
where \(V_{1}(\vec{x})\) is a volume centered at \(\vec{x}\), large compared to microscopic fluctuations but small compared to the system's size. For electromagnetic radiation, the relevant microscopic length scale is set by the wavelength. Since a cavity supports modes over a broad range of wavelengths, a natural choice is the wavelength at which the Planck spectrum reaches its maximum, as most of the radiative energy is concentrated near this wavelength. For solar radiation, the peak occurs at approximately \(\lambda_{\rm peak}\simeq 500\,\mathrm{nm}\), which corresponds to a characteristic averaging volume of order \((\lambda_{\rm peak})^3 \sim \mu\mathrm{m}^3\). In addition to spatial averaging, one may also perform averaging over a short time interval, chosen to be long compared to microscopic oscillations but short relative to macroscopic dynamics.

For concreteness, let us consider the contribution of the electric field to the energy density \eqref{energy-density-1}. We have
    \begin{align}
    \underline{u}_{E}(\vec{x},t) &:= \frac{1}{V_{1}(\vec{x})}\int_{V_{1}(\vec{x})}d^{3}x^{\prime}\left(\frac{1}{2}\epsilon_{0}\| \vec{E} (\vec{x}^{\prime},t)\|^{2}\right) \notag\\
    &= \frac{\epsilon_{0}}{2}\frac{1}{V_{1}(\vec{x})}\int_{V_{1}(\vec{x})}d^{3}x^{\prime}\int \frac{d^{3}k_{1}}{(2\pi)^{3}}\int \frac{d^{3}k_{2}}{(2\pi)^{3}} \notag\\
    &\qquad \Big[\left(\vec{E} (\vec{k}_{1})e^{i\vec{k}_{1}\cdot \vec{x}^{\prime}}e^{-i\omega (\vec{k}_{1})t} + \mathrm{c.c.}\right) \cdot \left(\vec{E} (\vec{k}_{2})e^{i\vec{k}_{2}\cdot \vec{x}^{\prime}}e^{-i\omega (\vec{k}_{2})t} + \mathrm{c.c.}\right)\Big]. 
    \end{align}
The spatial average over \(V_{1}(\vec{x})\) acts on the exponential factors \(e^{\pm i(\vec{k}_{1}\pm\vec{k}_{2})\cdot\vec{x}^{\prime}}\). For a volume large compared to the typical wavelength, these integrals become sharply peaked. More precisely,
    \begin{equation}\label{Dirac-approx}
    \frac{1}{V_{1}(\vec{x})}\int_{V_{1}(\vec{x})} d^{3}x^{\prime}\, e^{i(\vec{k}_{1}\pm\vec{k}_{2})\cdot\vec{x}^{\prime}} \approx \delta_{\vec{k}_{1},\mp\vec{k}_{2}} \quad\text{(discrete notation)},
    \end{equation}
or in the continuum limit it tends to a Dirac delta function \((2\pi)^{3}\delta^{3}(\vec{k}_{1}\pm\vec{k}_{2})\) times a factor that depends on the shape of the volume, but crucially it vanishes when \(\vec{k}_{1}\neq\mp\vec{k}_{2}\). Therefore, only terms with \(\vec{k}_{2} = \pm\vec{k}_{1}\) survive the average. However, cross terms such as \(\vec{E}(\vec{k})\cdot\vec{E}(-\vec{k})\) or its complex conjugate come with oscillatory factors \(e^{\pm 2i\omega(\vec{k})t}\) which, when averaged over a time interval long compared to the oscillation period, vanish. Hence, after performing both spatial and temporal averaging, only the terms with \(\vec{k}_{2} = \vec{k}_{1}\) and without \(e^{\pm 2i\omega t}\) survive. A detailed justification of why time averaging eliminates cross terms between different frequencies and why only the diagonal terms \(\vec{k}_2 = \vec{k}_1\) survive is given in Appendix~\ref{App-Time-Average}. These terms come from the product \(\vec{E}(\vec{k}_{1})e^{i\vec{k}_{1}\cdot\vec{x}'}\cdot\vec{E}(\vec{k}_{2})^{*}e^{-i\vec{k}_{2}\cdot\vec{x}'}\) with \(\vec{k}_{2}=\vec{k}_{1}\). The spatial average then gives a factor of unity, and the time average removes the oscillation. A similar analysis holds for the magnetic part (see Appendix~\ref{App-B-density}). Collecting the surviving contributions, one finds $\overline{\underline{u}}_B(\vec x)=\overline{\underline{u}}_E(\vec x)$\footnote{Underline and overline denote average over volume and average over time, respectively.}
    \bea
    \overline{\underline{u}}(\vec x) &:=&\overline{\underline{u}}_E(\vec x)+\overline{\underline{u}}_B(\vec x)\nn\\&=& 2\epsilon_0 \int \frac{d^{3}k}{(2\pi)^{3}} \, \vec{E}(\vec{k})^{*} \cdot \vec{E}(\vec{k}). 
    \label{energy-density}
    \eea

\subsection{Poynting vector and energy flux}

Now we turn to the Poynting vector \cite{McDonald}, 
    \be
    \vec{S} = \frac{1}{\mu_0} \, \vec{E} \times \vec{B},
    \label{Poynting-vector}
    \ee
which in turn, specifies the flux of energy. Using \eqref{E-solution} and \eqref{Fourier-magnetic-field},
    \begin{align}
    \vec S(\vec{x},t) &= \frac{1}{\mu_{0}} \frac{1}{V_{1}(\vec{x})}\int_{V_{1}(\vec{x})} d^{3}x^{\prime}\int \frac{d^{3}k_{1}}{(2\pi)^{3}}\frac{d^{3}k_{2}}{(2\pi)^{3}} \notag\\
    &\qquad \Bigg[ \left(\vec{E}(\vec{k}_{1})e^{i\vec{k}_{1}\cdot\vec{x}^{\prime}}e^{-i\omega(\vec{k}_{1})t} + \mathrm{c.c.}\right) \notag\\
    &\qquad \times \left( \frac{\vec{k}_{2}}{\omega(\vec{k}_{2})}\times \vec{E}(\vec{k}_{2})e^{i\vec{k}_{2}\cdot\vec{x}^{\prime}}e^{-i\omega(\vec{k}_{2})t} + \mathrm{c.c.}\right)\Bigg]. 
    \end{align}
As before, we need the average over a small spatial volume and a short time interval. Applying the same averaging procedure (the spatial average enforces \(\vec{k}_{1} = \pm\vec{k}_{2}\) and the time average kills the oscillatory terms) together with the Coulomb law \eqref{coulomb}, and the identity $\mu_0\epsilon_0=c^{-2}$ we obtain
    \begin{equation}
    \overline{\underline{\vec S}}(\vec{x}) =c\int \frac{d^{3}k}{(2\pi)^{3}} u(\vec{k}) \, \hat{k}, 
    \end{equation}
where  \(u(\vec{k}):=2\epsilon_0  \vec{E}(\vec{k})^{*} \cdot \vec{E}(\vec{k}), \) is the contribution of mode \(\vec{k}\) to the average energy density \eqref{energy-density}.

We assume that the radiation spectrum is isotropic \cite{Boyer2018,Boyer2021}, so that \(u(\vec{k})\) depends only on \(k = \|\vec{k}\|\). To find the energy escaping through a small hole, consider an infinitesimal surface element oriented perpendicular to the \(z\)-axis. Only modes with positive \(k_z\) contribute to the net outward flux. For each wave vector \(\vec{k}\), the component of the Poynting vector normal to the surface is \(c\,u(k)\cos\theta\). Integrating over the hemisphere \(\theta\in[0,\pi/2]\) gives
    \begin{equation}
    \label{I}
    I = c \int \frac{d^{3}k}{(2\pi)^{3}} u(k) \cos\theta = c \int \frac{dk}{(2\pi)^{3}} k^{2}u(k) \int_{0}^{2\pi}d\phi \int_{0}^{\pi/2} \cos\theta \sin\theta \, d\theta. 
    \end{equation}
The angular integral equals \(\pi\). The full energy density (integrated over all directions) is
    \begin{equation}
    \label{u}
    u = \int \frac{d^{3}k}{(2\pi)^{3}} u(k) = 4\pi \int \frac{dk}{(2\pi)^{3}} k^{2} u(k). 
    \end{equation}
Comparing \eqref{I} and \eqref{u} we recover \eqref{intensity}. 

Thus, by carefully separating modes according to their direction of propagation, we obtain the same relation \(I = \frac{c}{4}u\) that, in the photon picture, follows from kinetic theory. The advantage of the present derivation is that it uses only classical wave physics and explicit averaging, without any appeal to light quanta.
\subsection{Radiation pressure from the Maxwell stress tensor}\label{sec:tensor}

The momentum flux is described by the Maxwell stress tensor
\begin{equation}
T_{ij} = \epsilon_0\left(E_iE_j - \textstyle \frac{1}{2}\delta_{ij}\| \vec{E}\| ^2\right) + \frac{1}{\mu_0}\left(B_iB_j - \textstyle \frac{1}{2}\delta_{ij}\| \vec{B}\| ^2\right). 
\end{equation}
For a surface perpendicular to the \(i\)-direction, the pressure is \(P_i = -T_{ii}\). For example,
\begin{equation}
P_{x} = -\epsilon_{0}\Big(E_{x}^{2} - \textstyle \frac{1}{2}\| \vec{E}\| ^{2}\Big) - \frac{1}{\mu_{0}}\Big(B_{x}^{2} - \textstyle \frac{1}{2}\| \vec{B}\| ^{2}\Big). 
\label{pressure-x-component}
\end{equation}
In thermodynamic equilibrium, the radiation field is isotropic, so the pressure is the same in all directions. The diagonal components of the stress tensor therefore satisfy \(P_x = P_y = P_z\). Averaging over these components and using the fact that the trace of the Maxwell stress tensor equals \(-u\), we verify \eqref{pressure}. 

In the next section, we re-derive \eqref{intensity} and \eqref{pressure} by using a more formal ensemble method that does not rely on solving the equations of motion explicitly.

\section{A Systematic Field Theoretic Treatment}\label{Section-Abstract}\label{sec:formalism}

In contrast to the intuitive, wave-based derivation of Section~\ref{sec:intuitive}, we now develop a systematic statistical mechanical treatment of the classical electromagnetic field. Instead of solving Maxwell's equations and performing time/space averages, we work directly with the partition function and ensemble averages. This approach is more abstract, but it does not rely on the exact solvability of the equations of motion and naturally handles the constraints and the gauge invariance which characterise electrodynamics as a classical gauge field theory. The presentation is self-contained and follows the canonical Hamiltonian formulation, culminating in the derivation of \(P = u/3\) from the Maxwell stress tensor.

In the previous section, we tried to advance the discussion by appealing to intuition and to what is actually being measured. Statistical mechanics, however, has developed a framework that strengthens these intuitive methods with rigorous mathematics. In this section, therefore, we encounter more mathematical manipulations. The appeal of the problem lies in the fact that we are dealing with a system that, in essence, is a field theory. We aim to construct a partition function for this field theory and thereby derive its thermodynamic relations. As we shall see, this field theory contains twists and subtleties that resemble the challenges encountered in quantum electrodynamics. We believe it is worthwhile to confront these complexities here, where the physics of the story remains closer to intuition.  

Before turning to electrodynamics, let us briefly review the general procedure of statistical mechanics. 
Instead of time averages, we employ ensemble averages; that is, we sum over different configurations of the system. 
Statistical mechanics relies on the ergodic hypothesis, which assumes that temporal and ensemble averages coincide. 
Thus, rather than following the detailed time evolution, we obtain the desired results by considering all possible configurations.  

The first step is to identify the configurations of the system. 
In classical mechanics, these are specified by the generalized coordinates and their conjugate momenta, i.e., by the system's {location} in phase space. 
For example, a particle moving in one dimension is described by the pair $(q,p)$. 
The Hamiltonian $H$ of the system is then defined in terms of these variables. 

The next step is to construct the appropriate ensemble, such as the canonical ensemble. In this case, the probability of a microstate $(q,p)$ is proportional to the Boltzmann factor 
\(
e^{-\beta H(q,p)},
\)
where \(\beta := \frac{1}{k_B T},\) with $k_B$ being Boltzmann's constant (which relates energy to temperature), and $T$ denoting the absolute temperature \cite{Boltzmann}.
In this framework, the thermal average of any observable $F(q,p)$ is defined as
    \be
    \langle F \rangle := \frac{1}{Z} \int dq \, dp \; F(q,p) \, e^{-\beta H(q,p)},
    \label{Ensemble-Average}
    \ee
where $Z$ is the partition function,
\be
Z := \int dq \, dp \, e^{-\beta H(q,p)}.
\ee
Quantities of interest include averages such as $\langle q^2 \rangle$, $\langle p^2 \rangle$, or $\langle qp \rangle$.


Returning now to electrodynamics, the steps we  follow in a statistical approach are:
\begin{enumerate}
     \item Define the canonical variables.
    \item Write down the Hamiltonian of the system.
    \item Construct the Boltzmann distribution.
    \item Compute the desired quantities by averaging with respect to the Boltzmann weight.
\end{enumerate}

Electromagnetism, however, introduces an additional layer of complexity that makes the problem both more challenging and more intriguing. The fundamental dynamical variables of electrodynamics are the electromagnetic potentials $(\phi,\vec A)$, from which the fields are derived. Accordingly, the Lagrangian and Hamiltonian are formulated in terms of these potentials, and the equations of motion follow from their variation. The potentials also introduce the subtle issue of gauge freedom: distinct potential configurations may correspond to the same physical state. To avoid overcounting such gauge-equivalent configurations in the ensemble, the integration measure in \eqref{Ensemble-Average} must be defined with care.

Presenting all of these arguments simultaneously would obscure their logic, so we have arranged them into a sequence of subsections. Each subsection isolates a distinct component of the calculation, making its role and conclusion transparent.

\subsection{Maxwell's Canonical Equations}

In this subsection, we identify the canonical field variables and their conjugate momenta, i.e., the phase space variables of Maxwell theory, and show how the Hamiltonian density is expressed in terms of them.

Maxwell's equations can be derived from the principle of least action, $S = \int dt\, L$, where the Lagrangian is
\begin{equation}
\label{Lagrangian-space}
L = \frac{1}{2} \int d^3x \left( \epsilon_0 \vec{E}^2 - \frac{1}{\mu_0} \vec{B}^2 \right).
\end{equation}

The electric and magnetic fields can be expressed in terms of the potential components $A^\mu = (\phi, \vec{A})$, which serve as the canonical variables of the theory \cite{gauge}:
\begin{equation}
\vec{B} = \nabla \times \vec{A}, 
\qquad 
\vec{E} = -\nabla \phi - \partial_t \vec{A}.
\end{equation}

Defining the Hamiltonian and the partition function requires the conjugate momenta. By definition,
    \be
    \pi_{\mu} := \frac{\delta S}{\delta (\partial_t A^{\mu})}.
    \ee
{This implies that} the momentum conjugate to the vector potential is proportional to the electric field,
\begin{equation}
\label{momentum-electric-field}
\pi_i = -\epsilon_0 E_i,
\end{equation}
while that of the scalar potential vanishes,
\begin{equation}
\label{pi0}
\pi_0 = 0.
\end{equation}
Although this may seem unusual, its significance becomes clear as we proceed.

We now construct the Hamiltonian:
\bea
H &:=& \int d^3x \, \pi_\mu \partial_t A^\mu - L \nn\\
&=& \int d^3x \left[  \left( \frac{\vec{\pi}^2}{2\epsilon_0} + \frac{\vec{B}^2}{2\mu_0} \right) - \vec{\pi} \cdot \nabla \phi \right] \nn\\
&=& \int d^3x \left[  \left( \frac{\vec{\pi}^2}{2\epsilon_0} + \frac{\vec{B}^2}{2\mu_0} \right) + \phi \, \nabla \cdot \vec{\pi} \right].
\eea

In the last step, the boundary term from integration by parts has been discarded, since it does not affect the field dynamics.

After constructing the Hamiltonian, we verify that Hamilton's equations reproduce Maxwell's equations. For the scalar potential, Hamilton's equation gives
\[
0 = \partial_t \pi_0 = -\frac{\delta H}{\delta \phi} = -\nabla \cdot \vec{\pi} = \epsilon_0 \nabla \cdot \vec{E}.
\]
So Gauss's law in vacuum \(\nabla\cdot\vec{E}=0\) emerges as a secondary constraint enforced by the primary constraint \(\pi_0=0\) \cite{Thibes}. For the vector potential,
\[
\partial_t \vec{A} = \frac{\delta H}{\delta \vec{\pi}} = \frac{\vec{\pi}}{\epsilon_0} + \nabla \phi,
\]
which reproduces \(\vec{E} = -\nabla \phi - \partial_t \vec{A}\) via \(\vec{\pi} = -\epsilon_0 \vec{E}\). Finally, for the conjugate momentum,
\[
\partial_t \vec{\pi} = - \frac{\delta H}{\delta \vec{A}} = - \frac{1}{\mu_0} \nabla \times \vec{B},
\]
yielding Amp\`{e}re's law in vacuum \(\nabla \times \vec{B}=c^{-2} \partial_t \vec{E}\).

The Hamiltonian $H$ is independent of $\pi_0$, so Hamilton's equation for the scalar potential formally yields
\[
\partial_t \phi = \frac{\delta H}{\delta \pi_0} = 0.
\]
At first sight this seems to constrain $\phi$, but recall that the electromagnetic potentials admit gauge freedom: 
\[
A'^{\mu} := A^{\mu} + \partial^{\mu} \chi.
\]
This gives rise to the same electric and magnetic fields as $A^\mu$, regardless of the choice of $\chi$. Thus, the condition $\partial_t \phi = 0$ is not a physical requirement but a reflection of the gauge redundancy.  To make this explicit, we introduce the extended Hamiltonian
\[
H' := H + \int d^3x \, \pi_0 \, \xi,
\]
where $\xi$ is an arbitrary Lagrange multiplier. Its equation of motion enforces the primary constraint $\pi_0 = 0$, while simultaneously {setting} 
\[
\partial_t \phi = \xi.
\]
In this formulation, the time evolution of $\phi$ is governed by the undetermined function $\xi$, which reflects the gauge freedom of the theory. The physical dynamics remain unchanged: $H'$ still reproduces Amp\`{e}re's law and the other Maxwell equations. 

To simplify the analysis, we fix the gauge by choosing $\phi = 0$. With this choice, the Hamiltonian can be expressed entirely in terms of $\vec{A}$ and $\vec{E}$:
\begin{equation}
\label{Hamiltonian-Box-1}
H = \tfrac{1}{2} \int d^3x \left( \epsilon_0 \, \|\vec{E}(\vec{x})\|^2 + \tfrac{1}{\mu_0} \, \|\nabla \times \vec{A}(\vec{x})\|^2 \right).
\end{equation}
This form is particularly useful for computing statistical averages. Because the Hamiltonian is quadratic in the canonical variables, the Boltzmann weight takes the form of a Gaussian.

While encouraging, this also highlights a complication: the number of degrees of freedom is infinite. This was anticipated from the outset, since we are dealing with a field theory. At each spatial point $\vec{x}$, there are three dynamical variables $(A_1, A_2, A_3)$ together with their three conjugate momenta.  
Accordingly, the partition function takes the form of a functional integral:
\begin{equation}
Z = \int \mathcal{D}\vec{A}(\vec{x})\, \mathcal{D}\vec{E}(\vec{x}) 
\exp \left[ -\frac{\beta}{2} \int d^3x \left( 
\epsilon_0 \|\vec{E}(\vec{x})\|^2 + \frac{1}{\mu_0} \|\nabla \times \vec{A}(\vec{x})\|^2 
\right) \right].
\end{equation}
Here, e.g.,  $\mathcal{D}\vec{A}(\vec{x})$ denotes integration over all possible configurations of the field $\vec{A}(\vec{x})$, a notation standard in the context of functional (path) integrals.

Although the partition function involves an infinite number of degrees of freedom, its quadratic form makes it Gaussian. This observation allows us to evaluate it systematically by treating each Fourier mode as an independent harmonic oscillator.

To keep the progression clear, we proceed in stages:
\begin{enumerate}
\item \textbf{Fourier expansion:} Decompose $\vec{A}$ and $\vec{E}$ into spatial Fourier modes to isolate independent degrees of freedom mode by mode.
\item \textbf{Constraint handling:} Recognize that not all modes are physical since the gauge freedom and the constraints reduce the physical phase space, and rewrite the Hamiltonian and the partition function accordingly.
\item \textbf{Gaussian evaluation:} Once the Hamiltonian is cast into the appropriate quadratic form over the physical modes, perform the Gaussian integrals and extract the final expressions for the desired quantities.
\end{enumerate}

We implement these steps in the following subsections.

\subsection{Fourier Space and Constraints on Electric Fields}

In this subsection, we express the fields in terms of their Fourier components and verify that this transformation is canonical, i.e., it preserves Hamilton's equations. In addition, to implement the constraint $\nabla \cdot \vec{E} = 0$, we see that some modifications to the Hamiltonian are appropriate. 

We begin by writing the vector potential and electric field in Fourier space:
\bea
\label{Fourier-transform-A}
\vec{A}(\vec{x}) &=& \int \frac{d^3k}{(2\pi)^3} 
 \vec{A}(\vec{k})\, e^{i\vec{k} \cdot \vec{x}},\\
 \label{Fourier-transform-E}
\vec{E}(\vec{x}) &=& \int \frac{d^3k}{(2\pi)^3} 
\vec{E}(\vec{k})\, e^{i\vec{k} \cdot \vec{x}}.
\eea 
Because these fields are real, their Fourier components satisfy
\be
\vec{E}(\vec{k})^* = \vec{E}(-\vec{k}), 
\qquad 
\vec{A}(\vec{k})^* = \vec{A}(-\vec{k}).
\label{reality-field}
\ee

Although the Fourier expansion may look straightforward, it is crucial to verify that the transformation preserves the canonical structure of the theory. In other words, the Fourier transformation from the position-space variables to their momentum-space counterparts must be a canonical transformation, preserving the form of the equal-time Poisson brackets.

From \eqref{momentum-electric-field}, the fundamental brackets in position-space are
\be
\{ A_i(\vec{x}), E_j(\vec{y}) \} = -\,\epsilon_0^{-1}\,\delta_{ij}\,\delta^{(3)}(\vec{x}-\vec{y}).
\ee
For the Fourier components
\bea
\vec{A}(\vec{k}) &=& \int d^3x\, \vec{A}(\vec{x})\,e^{-i\vec{k}\cdot\vec{x}}, \\
\vec{E}(\vec{k}) &=& \int d^3x\, \vec{E}(\vec{x})\,e^{-i\vec{k}\cdot\vec{x}},
\eea
one obtains
\be
\label{Poisson-Fourier}
\{ A_i(\vec{k}), E_j(\vec{k}')^* \} 
= -\,(2\pi)^3\,\epsilon_0^{-1}\,\delta_{ij}\,\delta^{(3)}(\vec{k}-\vec{k}').
\ee
Hence the canonical commutation relations are preserved under Fourier transformation, demonstrating that the map 
\((A_i(\vec{x}),E_j(\vec{x})) \mapsto (A_i(\vec{k}),E_j(\vec{k}))\) 
is indeed canonical.\footnote{This conclusion can be further substantiated by rewriting the action in Fourier space; see Appendix~\ref{App-canonical}.}

In this way, we can now express the Hamiltonian in terms of the Fourier components, which is straightforward:
\bea
H &=& \frac{1}{2} \int d^3x \left( \epsilon_0 \|\vec{E}(\vec{x})\|^2 
+ \frac{1}{\mu_0} \|\vec{B}(\vec{x})\|^2 \right)\nn\\
\label{Hamiltonian-Fourier}
&=& \frac{1}{2} \int \frac{d^3k}{(2\pi)^3} \left[
\epsilon_0\, \vec{E}(\vec{k})^* \cdot \vec{E}(\vec{k}) 
+ \frac{1}{\mu_0}\, \left(\vec{k} \times \vec{A}(\vec{k})^*\right) \cdot \left(\vec{k} \times \vec{A}(\vec{k})\right) 
\right].
\eea

Accordingly, the functional integral can be written in terms of the Fourier variables. That is, we sum over $\vec{A}(\vec{k})$ and $\vec{E}(\vec{k})$ in the partition function 
instead of $\vec{A}(\vec{x})$ and $\vec{E}(\vec{x})$:
\be
Z = \int \mathcal{D}\vec{A}(\vec{k}) \, \mathcal{D}\vec{E}(\vec{k}) \, 
\exp\!\left\{ -\beta\, H[\vec{A}(\vec{k}), \vec{E}(\vec{k})] \right\}.
\label{partition}
\ee

The only subtlety in evaluating the partition function arises from the fact that the electric field is not an unconstrained dynamical variable. A naive integration over all field configurations would overcount unphysical states. In vacuum, Gauss's law requires that  $\nabla \cdot \vec{E}(\vec{x}) = 0$, so the functional integral must be restricted to divergence-free configurations. While implementing this constraint directly in the functional measure is possible, it is often more convenient to proceed indirectly. Specifically, we modify the Hamiltonian so that configurations violating Gauss's law are assigned zero statistical weight and therefore do not contribute to the partition function.

\begin{figure}
  \centering
  \includegraphics[width=0.2\textwidth]{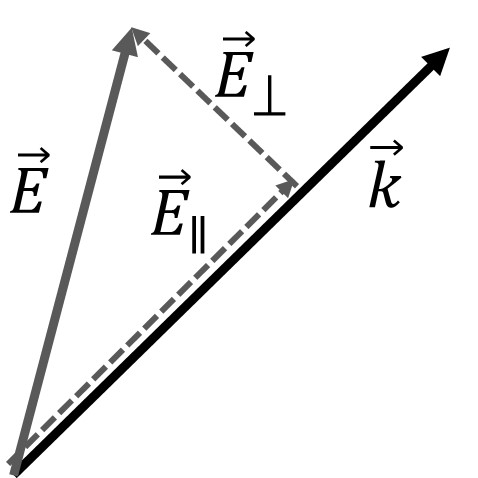}
  \caption{Decomposition of the field $\vec{E}$ into its longitudinal and transverse components with respect to the wave vector $\vec{k}$, namely $\vec{E}_{\parallel}$ and $\vec{E}_{\perp}$, respectively.}\label{Fig-3}
\end{figure}

In Fourier space, the constraint becomes
\(
\vec{k} \cdot \vec{E}(\vec{k}) = 0,
\)
which means that $\vec{E}(\vec{k})$ is perpendicular to $\vec{k}$. As shown in Figure~\ref{Fig-3}, the field $\vec{E}(\vec{k})$ may be uniquely decomposed into components parallel and perpendicular to the wave vector $\vec{k}$:
\be
\vec{E}(\vec{k}) = \vec{E}_\perp(\vec{k}) + \vec{E}_\parallel(\vec{k}).
\ee
If we can construct an operator that projects any $\vec{E}(\vec{k})$ onto its transverse component $\vec{E}_\perp(\vec{k})$, then only the allowed (divergence-free) part will contribute in the Hamiltonian, and the longitudinal part will be automatically discarded. We therefore look for an operator $\mathbf{P}(\vec{k})$ satisfying
\be
\mathbf{P}(\vec{k})\cdot\vec{E}(\vec{k}) = \vec{E}_\perp(\vec{k}).
\ee
Since $\vec{E}$ is a three-component vector field, $\mathbf{P}$ is a $3\times 3$ matrix that removes the component parallel to $\vec{k}$ and leaves the transverse components untouched. The projector is\footnote{It is straightforward to verify that $\mathbf{P}$ is idempotent, \(\mathbf{P}(\vec{k})^2 = \mathbf{P}(\vec{k})\). Moreover,  since $\mathbf{P}_{ij}(\vec k) k_j = 0$ the projector eliminates the longitudinal (unphysical) component of the field, leaving only the two transverse polarizations associated with physical electromagnetic waves.}
\be
\mathbf{P}_{ij}(\vec{k}) = \delta_{ij} - \frac{k_i k_j}{k^2}.
\ee

Consider the electric-field contribution to the Hamiltonian \eqref{Hamiltonian-Fourier}. 
To enforce the transversality condition, we replace that term by
\be
H_E = \frac{\epsilon_0}{2} \int \frac{d^3k}{(2\pi)^3} \,
\, \vec{E}(\vec{k})^* \cdot \mathbf{P}(\vec{k}) \cdot \vec{E}(\vec{k}),
\ee
where
\be
\vec{E}^* \cdot \mathbf{P} \cdot \vec{E} := \sum_{i,j} E_i^* \, P_{ij} \, E_j.
\ee
We may therefore integrate over all electric-field configurations in the partition function. In this formulation, the longitudinal components parallel to $\vec k$ carry no energy and hence enter the functional integral with unit weight, whereas the transverse components contribute exactly as in the original constrained theory.

It is important to note that the longitudinal modes, although energetically inert, generate an overall divergent multiplicative factor in the partition function $Z$, reflecting the infinite volume associated with the gauge redundancy. This divergence is, however, physically irrelevant. Observable quantities are computed as ensemble averages--ratios of functional integrals, as in \eqref{Ensemble-Average}--and the same divergent normalization factor appears in both the numerator and denominator. As a result, it cancels identically, and the unphysical degrees of freedom decouple from all physical observables. The remaining contributions arise solely from the transverse modes, yielding finite and well-defined results.

It is also straightforward to verify that
\be
\left(\vec{k} \times \vec{A}(\vec{k})^*\right) \cdot \left(\vec{k} \times \vec{A}(\vec{k})\right)
= k^2 \, \vec{A}(\vec{k})^* \cdot \mathbf{P}(\vec{k}) \cdot \vec{A}(\vec{k}),
\ee
so the projector $\mathbf{P}(\vec{k})$ arises naturally in the magnetic term as well, 
ensuring that $\vec{A}(\vec{k})$ has no longitudinal component. 
Thus the full Hamiltonian can be written compactly as
\be
\label{Hamiltonian-Box-2}
H = \frac{1}{2} \int \frac{d^3k}{(2\pi)^3} \left[
\epsilon_0 \, \vec{E}(\vec{k})^* \cdot \mathbf{P}(\vec{k}) \cdot \vec{E}(\vec{k})
+ \frac{k^2}{\mu_0} \, \vec{A}(\vec{k})^* \cdot \mathbf{P}(\vec{k}) \cdot \vec{A}(\vec{k})
\right].
\ee

\subsection{Calculation of field correlations using Gaussian integrals}

The derivation that follows is a standard exercise in the evaluation of quadratic (Gaussian) functional integrals. Readers interested only in the resulting correlation functions may safely skip ahead to Eqs. \eqref{correlation-E}, \eqref{correlation-A}, and \eqref{correlation-B}.

The partition function used to compute field correlations is based on the Hamiltonian \eqref{Hamiltonian-Box-2}, 
augmented by source terms linear in the fields. We write
    \be
    H_{\text{tot}} = H + H_J,
    \label{H-tot}
    \ee
with
\be
\label{Hamiltonian-physical}
H = \frac{\epsilon_0}{2} \int \frac{d^3k}{(2\pi)^3} \left[ 
 \vec{E}_\perp(\vec{k})^* \cdot \vec{E}_\perp(\vec{k}) 
+ c^2 k^2 \, \vec{A}_\perp(\vec{k})^* \cdot \vec{A}_\perp(\vec{k}) 
\right],
\ee
and
\be
H_J = \int \frac{d^3k}{(2\pi)^3} \Big(
E_\perp(\vec{k})^* \cdot J_E(\vec{k}) 
+ A_\perp(\vec{k})^* \cdot J_A(\vec{k})  \Big).
\ee
The source terms are introduced so that field correlations can be obtained by functional differentiation of the partition function.
Without loss of generality, they are supposed to be transverse, i.e.,
\begin{align}
\label{source}
    \mathbf{P}(\vec k)\cdot J_E(\vec k) &= J_E(\vec k), &
    \mathbf{P}(\vec k)\cdot J_A(\vec k) &= J_A(\vec k),
\end{align}
and satisfy the reality conditions
    \begin{align}
    &J_E(\vec k)^* =J_E(-\vec k), &J_A(\vec k) =J_A(-\vec k)^*.
    \label{reality-source}
    \end{align}

Henceforth, we  omit the subscript $\perp$ for convenience, with the understanding that $E(\vec{k})$ and $A(\vec{k})$ always denote transverse components. Following \eqref{Ensemble-Average} we define the correlator\footnote{Here and below, the functional measures \(\mathcal{D}E\) and \(\mathcal{D}A\) denote integrations over the transverse field configurations only, satisfying \(\vec{k}\cdot\vec{E}(\vec{k})=\vec{k}\cdot\vec{A}(\vec{k})=0\). The longitudinal components are not part of the functional integration.}
    \bea
    \langle E_i(\vec{k}) E_j^*(\vec{k}') \rangle 
    &:=& \frac{\int \mathcal{D}E \, \mathcal{D}A \; e^{-\beta H_{\text{tot}}} \, E_i(\vec{k}) E_j^*(\vec{k}')}
    {\int \mathcal{D}E \, \mathcal{D}A \; e^{-\beta H_{\text{tot}}}} \nn\\
    &=& \left. \frac{(2\pi)^6}{\beta^2 Z } 
    \frac{\delta^2 Z}{\delta [J_{E}(\vec{k})^*]_i \, \delta [J_{E}(\vec{k}')]_j} 
    \right|_{J_E = J_A = J_E^* = J_A^* = 0},
    \label{EE}
    \eea
where we have adopted the convention
\be 
\frac{\delta [J_E(\vec k)]_i}{\delta [J_E(\vec k')]_j}=\delta_{ij}\,\delta^3(\vec k-\vec k'),
\ee
used \eqref{reality-field} and \eqref{reality-source}, and noted that 
\be
H_J = \int \frac{d^3k}{(2\pi)^3} \Big(E(\vec{k}) \cdot J_E(-\vec{k}) + A(\vec{k}) \cdot J_A(-\vec{k}) \Big).
\ee
Similarly,
    \bea
    \langle A_i(\vec{k}) A_j^*(\vec{k}') \rangle 
    &=& \frac{\int \mathcal{D}E \, \mathcal{D}A \; e^{-\beta H_{\text{tot}}} \, A_i(\vec{k}) A_j^*(\vec{k}')}
    {\int \mathcal{D}E \, \mathcal{D}A \; e^{-\beta H_{\text{tot}}}} \nn\\
    &=& \left. \frac{(2\pi)^6}{\beta^2 Z } 
    \frac{\delta^2 Z}{\delta [J_{A}(\vec{k})^*]_i \, \delta [J_{A}(\vec{k}')]_j} 
    \right|_{J_E = J_A = J_E^* = J_A^* = 0}.
    \label{AA}
    \eea

To evaluate the Gaussian integrals, it suffices to complete the square in the exponential.\footnote{See Appendix \ref{App-square} for details.}
Working in Fourier space is advantageous because different wave numbers decouple and can be treated independently. 
The result is
\be
Z \propto \exp \left( \frac{\beta}{2\epsilon_0} \int \frac{d^3k}{(2\pi)^3} \left[ 
J_E^*(\vec{k}) \cdot \mathbf{P}(\vec{k}) \cdot J_E(\vec{k})
+ \frac{1}{c^2 k^2} J_A^*(\vec{k}) \cdot \mathbf{P}(\vec{k}) \cdot J_A(\vec{k}) 
\right] \right),
\label{partition-final}
\ee
where, following \eqref{source}, the projection operator $\mathbf{P}(\vec{k})$ is explicitly displayed for clarity. Since the proportionality constant contains no current-dependent terms, it is omitted. Using \eqref{partition-final} in \eqref{EE} and \eqref{AA}, we obtain
\bea
\label{correlation-E}
\langle E_i(\vec{k}) E_j^*(\vec{k}') \rangle 
&=& \frac{(2\pi)^3 k_B T}{\epsilon_0} \, \delta^3(\vec{k} - \vec{k}') \, P_{ij}(\vec{k}),\\ 
\label{correlation-A}
\langle A_i(\vec{k}) A_j^*(\vec{k}') \rangle 
&=& \frac{(2\pi)^3 \mu_0 k_B T}{k^2} \, \delta^3(\vec{k} - \vec{k}') \, P_{ij}(\vec{k}),
\eea
where the relation $\beta^{-1} = k_B T$ has been applied. Mixed correlations such as $\langle E_i A_j \rangle$ vanish identically, since the block-diagonal structure of the Hamiltonian forces the corresponding source terms to decouple.

Using the correlation functions~\eqref{correlation-E} and \eqref{correlation-A},  we can evaluate the thermal average of the Hamiltonian~\eqref{Hamiltonian-physical}. The result is
\be
\langle H \rangle = V \mathcal{N} \,(2 k_B T),
\label{Hamiltonian-Average}
\ee
where $V$ denotes the cavity volume,
\be
V :=\int d^3x = (2\pi)^3 \,\delta^3(\vec{k}-\vec{k}\,')\Big|_{\vec{k}'=\vec{k}},
\ee
and $\mathcal{N}$ is the total number of allowed wave--vectors
\be
\label{number}
\mathcal{N} := \int \frac{d^3k}{(2\pi)^3}.
\ee

Equation~\eqref{Hamiltonian-Average} shows that the thermal average of the Hamiltonian coincides with the familiar equipartition result: each wave--vector contributes the energy of two simple harmonic oscillators per unit volume. In the continuum description, however, both $V$ and $\mathcal{N}$ are apparently  infinite. The volume $V$ arises from the normalization of the Dirac delta function,
\[
(2\pi)^3\delta^3(\vec 0),
\]
corresponding to the infinite-volume limit, whereas $\mathcal{N}$ diverges because the continuum theory contains infinitely many electromagnetic modes with arbitrarily large wave vectors. Accordingly, both quantities should be understood as regulated by finite infrared and ultraviolet cutoffs, respectively. We retain the compact notation above for simplicity and return to the role of these regulators when discussing the energy density and radiation pressure later in this section.

The correlation of the magnetic fields follows directly from the correlator \eqref{correlation-A}. Since
\be
\vec{B} = \nabla \times \vec{A}, 
\qquad 
\vec{B}(\vec{k}) = i\,\vec{k}\times \vec{A}(\vec{k}).
\ee
we have
\bea
\langle B_i(\vec{k}) B_j^*(\vec{k}') \rangle 
&=& \frac{(2\pi)^3 \mu_0 k_B T}{k^2} \,
\epsilon_{imr}\,\epsilon_{jns}\,k_m k_n \, P_{rs}(\vec{k}) \,
\delta^3(\vec{k} - \vec{k}') \nn\\
\label{correlation-B}
&=& (2\pi)^3 \mu_0 k_B T \, P_{ij}(\vec{k}) \,
\delta^3(\vec{k} - \vec{k}'),
\eea
where $\epsilon_{imr}$ is the Levi-Civita symbol. 
In reaching the final form, we have used
\be
\epsilon_{imr}\,k_m k_r = 0,
\qquad
\epsilon_{imr}\,\epsilon_{jns}\,\delta_{rs} 
= \delta_{ij}\,\delta_{mn} - \delta_{in}\,\delta_{mj}.
\ee
This result is manifestly consistent with $\nabla \cdot \vec{B}(\vec{x}) = 0$, or equivalently $\vec{k}\cdot \vec{B}(\vec{k}) = 0$.

\noindent\textbf{Remark.} The transverse projector $P_{ij}(\vec{k})$ appearing in all correlators reflects the fact that only physical, divergence-free modes contribute. This is the statistical mechanics analogue of gauge invariance in the canonical formalism.

The correlations obtained in this {section} are very similar to the relations \eqref{correlation-E-rough} and \eqref{correlation-B-rough}, although they have been derived in a different way. In obtaining \eqref{correlation-E-rough} and \eqref{correlation-B-rough}, we use the equations of motion and performed averaging over time or within a finite volume. Here, by contrast, we carried out ensemble averaging, and the two approaches coincide. According to the ergodic theorem, time (or spatial) averages and ensemble averages are equivalent.
Furthermore, the transverse projector $\mathbf{P}_{ij}$ appears explicitly in \eqref{correlation-E} and \eqref{correlation-B}, whereas in \eqref{correlation-E-rough} and \eqref{correlation-B-rough} it does not. The distinction stems from the fact that the constraints are already built into the solutions of the equations of motion. Consequently, the resulting fields are automatically transverse, and the projector need not be written explicitly. In the statistical approach, however, one integrates over arbitrary field configurations, so the constraints must be imposed explicitly.

Another distinction concerns the constants multiplying the Dirac delta functions. In the statistical derivation presented here, these coefficients appear naturally. In the former approach, they would have to be obtained by invoking the equipartition theorem, which assigns an average energy $k_B T$ to each oscillating mode. This yields coefficients consistent with those derived in the present framework.

\subsection{Energy density and radiant power}

We now return to our central question: the relationship between energy density and radiant power. Within the present statistical framework, this connection cannot be established directly. While the electromagnetic energy density can be computed in this way (albeit with a divergent result), the radiated power requires analysis of the Poynting vector \eqref{Poynting-vector}, and, in particular, its average value. Because the correlations between electric and magnetic fields vanish,
\be
\langle E_i(\vec{k}) B_j(\vec{k}) \rangle = 0,
\ee
the average Poynting vector in the thermal equilibrium is identically zero. This vanishing of the average Poynting vector is the correct equilibrium result (no net energy flow). It does not contradict the relation \(I = \frac{c}{4}u\) derived in Section~\ref{sec:intuitive}, because there we specifically isolated modes moving toward a hole, i.e., integrated only over a hemisphere in \(k\)-space. This outcome is consistent with physical intuition: a nonzero average flux would imply a net transfer of energy, contradicting the very notion of equilibrium.

In section \ref{sec:intuitive}, by solving the equations of motion, we were able to identify the propagation direction of individual waves. At a given point, left-moving and right-moving modes could be distinguished, and the flux associated with each category computed. Such a decomposition, however, is not available in the ensemble approach, which yields only the equilibrium result.

Nevertheless, these calculations remain highly instructive. First, they sharpen our intuition about electromagnetic systems in {thermal} equilibrium and, more broadly, about the principles of statistical mechanics. Second, they provide a natural entry point into field-theoretic methods and path integrals, which play a central role in quantum electrodynamics. Finally, they enable estimation of radiation pressure -- a quantity of fundamental importance that, as discussed earlier, leads directly to the Stefan-Boltzmann law.

To compute the thermal average of the radiant pressure, we note that, as already explained in subsection \ref{sec:tensor}, the pressure can be obtained from the stress tensor. While this calculation can be carried out in Fourier space, here we present the derivation directly in position space. Using \eqref{Fourier-transform-E} together with \eqref{correlation-E}, we find
\be
\left\langle E_i(\vec{x}) E_j(\vec{x}\,') \right\rangle 
= \frac{ k_B T}{\epsilon_0} \int \frac{d^3k}{(2\pi)^3} \, P_{ij}(\vec{k}) \, e^{i \vec{k} \cdot (\vec{x} - \vec{x}')}.
\ee
For the correlation at coincident points, $\vec{x}=\vec{x}\,'$, this reduces to\footnote{Details are given in Appendix~\ref{App-evaluate}.}
\be
\label{correlator-E-space}
\left\langle E_i(\vec{x}) E_j(\vec{x}) \right\rangle 
= \frac{k_B T}{\epsilon_0} \int \frac{d^3k}{(2\pi)^3} P_{ij}(\vec{k}) 
= \frac{2}{3} \frac{k_B T}{\epsilon_0} \, \mathcal{N}\, \delta_{ij},
\ee
where \(\mathcal{N}\) is defined in \eqref{number}. This result, in particular, shows that
\be
\langle E_i(\vec{x})^2 \rangle 
= \frac{1}{3} \langle \|\vec{E}(\vec{x})\|^2 \rangle.
\ee
An analogous relation holds for the magnetic field. Substituting into the pressure expression \eqref{pressure-x-component} and performing the ensemble average, we arrive at
\be
\langle P\rangle =  \epsilon_0 \frac{\langle \|\vec{E}(\vec{x})\|^2 \rangle}{6} 
+ \frac{\langle \|\vec{B}(\vec{x})\|^2 \rangle}{6\mu_0} 
= \frac{\langle u\rangle}{3},
\ee
in agreement with \eqref{pressure}. 

Before closing this section, let us emphasize that \eqref{correlator-E-space} is fully consistent with the equipartition theorem. In particular, it yields
\be
\frac{\epsilon_0}{2}\,\langle \|\vec{E}(\vec{x})\|^2 \rangle = \mathcal{N} k_B T,
\ee
which expresses the expected thermal contribution of each mode to the electric field energy.

  At this point, it is appropriate to comment on the formal divergences encountered above. In the continuum limit, the mode count $\mathcal{N}$ diverges because of the unlimited number of electromagnetic modes with arbitrarily large wave vectors. Introducing a finite ultraviolet cutoff renders the energy density and the coincident-point correlation functions finite, while the infinite-volume normalization is handled by the usual infrared regularization. Since both the pressure and the energy density are proportional to the same regulated quantity $\mathcal{N}$, the regulator dependence cancels in their ratio. Consequently, the equation of state
\(
P=\frac{u}{3}
\)
is independent of the cutoff, even though the individual quantities diverge as the cutoff is removed.

To summarise, classical statistical mechanics correctly reproduces the universal proportionalities
\[
P=\frac{u}{3}, \qquad
I=\frac{c}{4}u,
\]
the latter having been derived in Section~\ref{sec:intuitive}. Together with thermodynamics, these relations imply the Stefan--Boltzmann law $u\propto T^4$. What classical theory fails to determine is the finite value of the proportionality constant: because of the ultraviolet catastrophe, the energy density itself diverges as the ultraviolet cutoff is removed. Nevertheless, the present analysis remains valuable both as a classical derivation of these universal relations and as a natural bridge to finite-temperature quantum field theory.

\section{Conclusion}
\label{sec:conclusion}

In this paper we have revisited the classical foundations of the statistical mechanics of electromagnetic radiation. Starting from Maxwell's equations and classical thermodynamics, we have shown that two fundamental proportionalities -- the energy flux \(I = \frac{c}{4}u\) and the radiation pressure \(P = \frac{u}{3}\) -- follow without any appeal to photons or quantum statistics. The former was obtained through an intuitive wave-based derivation that uses spatial and temporal averaging and isolates modes propagating toward a hole; the latter was re-derived in a systematic, constraint based field-theoretic approach that explicitly handles gauge invariance via a transverse projector and evaluates thermal averages from the classical partition function.

These derivations make it clear that while the classical theory is unable to produce the correct blackbody spectrum (it suffers from the ultraviolet catastrophe and gives a divergent energy density) and therefore cannot fix the absolute constant in the Stefan-Boltzmann law \(u = A T^{4}\), the functional relations among energy density, flux, and pressure are already contained in classical electromagnetism and classical statistical mechanics. In this sense, the failure of the classical theory is not total: it correctly predicts the structure of the macroscopic laws, even though it fails at the level of the spectral distribution.

The formulation presented here addresses a gap often overlooked in the standard literature. It demonstrates how to systematically construct the partition function for a classical gauge field theory, enforce gauge constraints, and evaluate Gaussian functional integrals and thermal averages within a concrete physical system. This framework establishes a clear classical foundation for more advanced topics, such as finite-temperature quantum field theory and path-integral methods. The two complementary approaches—the heuristic averaging method and the systematic ensemble method—clarify the interplay between classical fields, gauge invariance, and thermodynamics.

\section*{Appendices}
\appendix

\section{Time Averaging of Rapid Oscillations}
\label{App-Time-Average}

Physical measurements are always performed over a finite time interval, which naturally introduces a time average. This averaging plays a role analogous to spatial averaging and further justifies the approximations used in Section~\ref{sec:intuitive}. To illustrate the effect, consider a simple one-dimensional example: the time-averaged energy density of a superposition of two monochromatic waves. The (complex) electric field is
\begin{equation}
E(x,t) = E_{1}e^{i(k_{1}x - \omega_{1}t)} + E_{2}e^{i(k_{2}x - \omega_{2}t)}. 
\end{equation}
Computing the energy density requires squaring the field, which produces two types of contributions: (i) terms from the square of each individual wave, and (ii) cross terms arising from the product of the two waves. We focus on the cross term, denoted \(u_{12}\). Its time average over an interval of duration \(T\) is
\begin{equation}
\bar u_{12} := \frac{1}{T}\int_{-T/2}^{T/2}dt\, u_{12}(t) = \frac{1}{T} E_{1}E_{2}^{*}e^{i(k_{1} - k_{2})x}\int_{-T/2}^{T/2}dt\, e^{i(\omega_{1} - \omega_{2})t}+\mathrm{c.c.}.
\end{equation}
Evaluating the integral gives
\begin{equation}
\bar u_{12} \propto \frac{2}{(\omega_{1} - \omega_{2})T}\sin\big[(\omega_{1} - \omega_{2})T\big]. 
\end{equation}
The sine factor is bounded between \(-1\) and \(1\), but the prefactor can be extremely small. For example, even the frequency difference between yellow and green light is of order \(10^{15}\,\text{rad/s}\). With a measurement time of the order of seconds, the resulting contribution is suppressed to about \(10^{-15}\) -- effectively zero. Only if the frequency difference were as small as \(2\pi\) radians per second would the integral yield a contribution of order unity. For optical frequencies, this would require a measurement apparatus capable of resolving frequencies to fifteen significant digits.

Thus, time averaging eliminates cross terms between waves of different frequencies, leaving only the incoherent sum of their individual energy densities. The net effect is that only the energies of the individual modes survive, and the total energy density is obtained by summing over them. The magnetic-field contribution follows by the same reasoning.

More formally, the averaging procedure can be expressed through correlation functions,
\begin{align}
\label{correlation-E-rough}
&\langle \vec{E}(\vec{k}, t) \cdot \vec{E}(\vec{k}', t) \rangle_{t,V} =0,&\langle \vec{E}(\vec{k}, t) \cdot \vec{E}(\vec{k}', t)^* \rangle_{t,V} = C_E \, \delta(\vec{k} - \vec{k}'),
\\
\label{correlation-B-rough}
&\langle \vec{B}(\vec{k}, t) \cdot \vec{B}(\vec{k}', t) \rangle_{t,V} =0&\langle \vec{B}(\vec{k}, t) \cdot \vec{B}(\vec{k}', t)^* \rangle_{t,V} = C_B \, \delta(\vec{k} - \vec{k}'),
\end{align}
where \(C_E\) and \(C_B\) are constants.These relations emerge after solving Maxwell's equations and subsequently performing the time and spatial averages.  This justifies the step in Section~\ref{sec:intuitive} where we dropped cross terms between different Fourier modes. In Section~\ref{Section-Abstract}, however, we demonstrate how such expressions can be derived directly from statistical mechanics, without recourse to the explicit solution of the field equations.

\section{Magnetic Contribution to the Energy Density}\label{App-B-density}

Substituting the Fourier representation of the magnetic field \eqref{Fourier-magnetic-field} into the definition of the energy density \eqref{energy-density-1}, we obtain
\begin{align}
\underline{u}_B(\vec{x}, t) 
= \frac{1}{2\mu_0}\frac{1}{V_1(\vec x)}  
\int_{V_1(\vec{x})} d^3x' 
\int \frac{d^3k_1}{(2\pi)^3} \int \frac{d^3k_2}{(2\pi)^3}
\Bigg[&\left( \frac{\vec k_1}{\omega(\vec k_1)}\times \vec{E}(\vec{k}_1)\, e^{i\vec{k}_1 \cdot \vec{x}'} e^{-i\omega(\vec{k}_1)t} + \text{c.c.} \right) \nn\\
&\cdot \frac{\vec k_2}{\omega(\vec k_2)}\times \vec{E}(\vec{k}_2)\, e^{i\vec{k}_2 \cdot \vec{x}'} e^{-i\omega(\vec{k}_2)t}\Bigg] + \text{c.c.} .
\end{align}
Applying the approximation \eqref{Dirac-approx}, this expression simplifies to
    \begin{align}
    \underline{u}_B(\vec{x}, t) 
    = \frac{1}{2\mu_0}  
    \int \frac{d^3k}{(2\pi)^3} 
    \left(- \frac{\vec k\times  \vec{E}(\vec{k})\cdot \vec k\times  \vec{E}(-\vec{k})}{\omega(\vec k)^2}\, e^{-2i\omega(\vec{k})t} 
    + \frac{\vec k\times  \vec{E}(\vec{k})^*\cdot \vec k\times  \vec{E}(\vec{k})}{\omega(\vec k)^2}\right) + \text{c.c.} ,
    \end{align}
and averaging over a suitable time period gives 
    \begin{align}
    \overline{\underline{u}}_B(\vec{x}) = \frac{1}{\mu_0}  \int \frac{d^3k}{(2\pi)^3} \frac{\vec k\times  \vec{E}(\vec{k})^*\cdot \vec k\times  \vec{E}(\vec{k})}{\omega(\vec k)^2}. 
    \end{align}
Using \eqref{coulomb},   the dispersion relation $\omega(\vec k)^2=c^2k^2$, and the identity $c^{-2}=\epsilon_0\mu_0$ we obtain
\begin{equation}
\label{B-density}
\overline{\underline{u}}_B(\vec{x}) = \epsilon_0 \int \frac{d^3k}{(2\pi)^3} \, \vec{E}(\vec{k})\cdot \vec{E}(\vec{k})^* .
\end{equation}

\section{Canonical Structure of the Fourier Transformation}\label{App-canonical}

In this appendix we verify that the Fourier transformation relating the position-space and momentum-space canonical variables is a  \emph{canonical transformation}.

Using \eqref{Fourier-transform-A} and \eqref{Fourier-transform-E}, the quadratic terms of the Lagrangian \eqref{Lagrangian-space} can be written in Fourier space as
\bea
\int d^3x\,\|\vec{E}(\vec{x})\|^2 
&=&  \int \frac{d^3k}{(2\pi)^3}\,\vec{E}(\vec{k})\cdot\vec{E}^*(\vec{k}), \\
\int d^3x\,\|\nabla\times\vec{A}(\vec{x})\|^2 
&=&  \int \frac{d^3k}{(2\pi)^3}\,\big(\vec{k}\times\vec{A}(\vec{k})\big)\cdot\big(\vec{k}\times\vec{A}^*(\vec{k})\big).
\eea

Since the relation between the electric field and the potentials is linear, their Fourier transforms are likewise linearly related:
\be
\vec{E}(\vec{k}) = -\partial_t \vec{A}(\vec{k}) - i\vec{k}\,\phi(\vec{k}).
\ee
Substituting this into the Lagrangian \eqref{Lagrangian-space} yields
\bea
L = \frac{1}{2} \int \frac{d^3k}{(2\pi)^3} \Big[ \epsilon_0
\big(\partial_t \vec{A}(\vec{k}) + i\vec{k}\,\phi(\vec{k})\big)\cdot
\big(\partial_t \vec{A}^*(\vec{k}) - i\vec{k}\,\phi^*(\vec{k})\big) 
- \big(\vec{k}\times\vec{A}(\vec{k})\big)\cdot\big(\vec{k}\times\vec{A}^*(\vec{k})\big)
\Big].
\eea
From this expression the canonical momentum conjugate to $\vec{A}(\vec{k})$ follows directly:
\be
\vec{\pi}_{\vec{A}(\vec{k})} := \frac{\delta L}{\delta \partial_t \vec{A}(\vec{k})}
= \frac{\epsilon_0}{(2\pi)^3}\big(\partial_t \vec{A}^*(\vec{k}) - i\vec{k}\,\phi^*(\vec{k})\big)
= -\,\frac{\epsilon_0}{(2\pi)^3}\,\vec{E}^*(\vec{k}).
\ee

Thus the conjugate momentum to $\vec{A}(\vec{k})$ is precisely \(-\,\tfrac{\epsilon_0}{(2\pi)^3}\,\vec{E}^*(\vec{k})\), in agreement with the Poisson bracket \eqref{Poisson-Fourier}. This establishes that the Fourier transformation is not merely a convenient change of variables, but a genuine \emph{canonical transformation}.
\section{Completing the square in the exponential}\label{App-square}

The contribution of the electric field to the total Hamiltonian $H_{\rm tot}$ \eqref{H-tot} can be written as     
\bea 
\mathcal{H}_E&:=&\int \frac{d^3k}{(2\pi)^3}\left[\frac{\epsilon_0}{2}\vec E(\vec k)\cdot\vec E(-\vec k)+\vec E(\vec k)\cdot\vec J_E(-\vec k)\right]\nn\\
&=&\frac{\epsilon_0}{2}\int \frac{d^3k}{(2\pi)^3}\left[\left(\vec E(-\vec k)+\frac{1}{\epsilon_0}\vec J_E(-\vec k)\right)\cdot \left(\vec E(\vec k)+\frac{1}{\epsilon_0}\vec J_E(\vec k)\right)\right]\nn\\&-&
\frac{1}{2\epsilon_0}\int \frac{d^3k}{(2\pi)^3} \vec J_E(-\vec k)\cdot \vec J_E(\vec k).
\eea
\section{Evaluating the integral \eqref{correlator-E-space}}\label{App-evaluate}
We expand in spherical coordinates:
\be
\int \frac{d^3k}{(2\pi)^3} P_{ij}(\vec{k}) 
= \left( \int \frac{dk}{(2\pi)^3} \, k^2 \right) 
\int d\Omega \left( \delta_{ij} - \hat{k}_i \hat{k}_j \right).
\ee
The angular and radial parts separate, yielding
\be
\int d\Omega \left( \delta_{ij} - \hat{k}_i \hat{k}_j \right) 
= \frac{8\pi}{3} \delta_{ij},
\ee
and
\be
\int \frac{dk}{(2\pi)^3} k^2 
= \frac{1}{4\pi} \int \frac{d^3k}{(2\pi)^3} 
= \frac{1}{4\pi} \mathcal{N}.
\ee

\end{document}